\documentclass[prb,twocolumn, superscriptaddress,floatfix]{revtex4}
\usepackage{graphicx}
\usepackage{epstopdf}
\usepackage{bm, amsmath, amssymb}
\usepackage{pdfpages} 

\newcommand{\nbar}[0]{\bar{n}}

\newcommand{\adag}[0]{a^\dagger}
\newcommand{\omegr}[0]{\omega_0}

\begin{document}

\title{Quantum State Sensitivity of an Autoresonant Superconducting Circuit}

\author{K. W. Murch}
\affiliation{Quantum Nanoelectronics Laboratory, Department of Physics, University of California, Berkeley CA 94720}
\author{E. Ginossar}
\affiliation{Advanced Technology Institute and Department of Physics, University of Surrey, Guildford, GU2 7XH, United Kingdom}
\author{S. J. Weber}
\affiliation{Quantum Nanoelectronics Laboratory, Department of Physics, University of California, Berkeley CA 94720}
\author{R. Vijay}
\affiliation{Quantum Nanoelectronics Laboratory, Department of Physics, University of California, Berkeley CA 94720}
\author{S.M. Girvin}
\affiliation{Department of  Physics, Yale University, New Haven, CT 06520-8120}
\author{I. Siddiqi}
\affiliation{Quantum Nanoelectronics Laboratory, Department of Physics, University of California, Berkeley CA 94720}

\date{\today}

\begin{abstract}
When a frequency chirped excitation is applied to a classical high-Q nonlinear oscillator, its motion becomes dynamically synchronized to the drive and large oscillation amplitude is observed, provided the drive strength exceeds the critical threshold for autoresonance. We demonstrate that when such an oscillator is strongly coupled to a quantized superconducting qubit, both the effective nonlinearity and the threshold become a non-trivial function of the qubit-oscillator detuning. Moreover, the autoresonant threshold is sensitive to the quantum state of the qubit and may be used to realize a high fidelity, latching readout whose speed is not limited by the oscillator Q. \end{abstract}

\maketitle

In a nonlinear oscillator, the resonance frequency shifts as a function of the oscillation amplitude. When such a system is driven, high oscillation energy can be attained if the frequency of the excitation is modulated to maintain resonance. Remarkably, complete synchronization between the forcing field and the oscillator motion can be achieved by using a {\em sufficiently} strong drive whose frequency is linearly modulated (chirped) in time. If the drive strength, however, is weaker than a critical value, then the system does not synchronize and the forced oscillations decay. This phenomenon, known as autoresonance (AR), plays an important role in diverse fields of classical physics ranging from planetary dynamics\cite{malh93pluto} to plasma physics\cite{faja01ajp}, and has recently been demonstrated at microwave frequencies in superconducting Josephson junction circuits \cite{naam08chirp}. Further work has explored the interplay between autoresonance and quantum mechanics by cooling a weakly anharmonic circuit such that its ground state is dominated by quantum fluctuations\cite{murc11chirp} and by realizing a circuit with a highly anharmonic level structure \cite{shal12}. In the former, quantum fluctuations simply broaden the classical AR threshold \cite{murc11chirp, bart11} whereas in the latter, the dynamics are described by quantum ladder climbing \cite{marc04ladder}. 

In this letter we report the observation of AR in a superconducting circuit where a weakly nonlinear, planar microwave cavity is strongly coupled to a transmon qubit in a circuit quantum electrodynamics architecture. We observe that both the value and width of the critical threshold for AR depend sensitively on the quantum level structure of the qubit-cavity system, exhibiting a rich dependence on qubit-cavity detuning. Our analysis points to the conclusion that the initial quantum dynamics which occurs in the few lowest states of the system crucially influences the final outcome of an applied chirp excitation, forming the basis of a sensitive qubit state readout. Furthermore, we show that analogous to the classical AR threshold, the variation of qubit state sensitivity with detuning is correlated with the low-power effective nonlinearity of the system. At points of optimal bias, we achieve high readout fidelity and demonstrate that a qubit coupled to a high Q cavity can be effectively measured in a chirp time that is significantly shorter than the cavity equilibration time.

   \begin{figure}
\includegraphics[angle = 0, width = 0.5\textwidth]{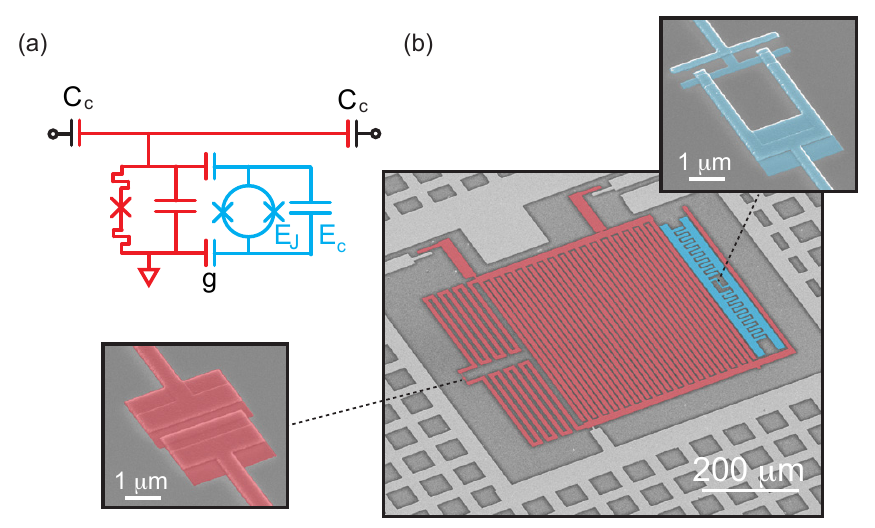}
\caption{\label{fig3} Experimental device: a transmon qubit coupled to a nonlinear superconducting cavity.  (a) Circuit diagram of the device; the anharmonic resonator is formed from a meander inductor embedded with a Josephson junction and an interdigitated capacitor.  The resonator is isolated from the 50 $\Omega$ environment by coupling capacitors $C_c$ and coupled to a transmon qubit characterized by Josephson and charging energy scales $E_\mathrm{J}$ and $E_\mathrm{c}$ respectively. The coupling rate is $g$.  (b) Scanning electron micrograph of the the device with the resonator and qubit junctions (lower and upper insets).}
\end{figure}

Our device, shown in Figure 1,  consists of two superconducting, lumped element planar non-linear resonators that are cooled to their quantum ground state.  Though both oscillators consist of capacitively shunted Josephson junctions, the parameters are chosen such that one plays the role of a quasi-classical resonator and the other as a multilevel artificial atom.  The quasi-classical resonator is composed of a 1.8 $\mu$A Josephson tunnel junction embedded in $1.4$ nH of linear inductance, and shunted with $580$ fF of capacitance to realize a resonance at $\omegr/2\pi =  5.3445$ GHz.  The total $\mathrm{Q}=9 \times 10^3$ of the resonator is limited by deliberate coupling to the environment as set by coupling capacitors $C_\mathrm{c}$, and by losses in the Josephson junction. The transmon qubit\cite{koch07transmon} with Josephson and charging energies $E_\mathrm{J}= 100$ GHz and $E_\mathrm{C} = 280$ MHz, respectively, is composed of an interdigitated capacitor shunted with a two junction SQUID with $100$ nA junctions.  The qubit $0\rightarrow 1$ transition frequency can be tuned between $3$ and $15$ GHz by adjusting an applied flux through the SQUID loop. The qubit--resonator coupling rate is $g/2\pi =118$ MHz.  The device was fabricated using standard e-beam lithography and angle deposition of aluminum on high resistivity silicon \cite{murc11chirp}.  The chip was thermally anchored to the mixing chamber stage of a dilution refrigerator ($T = 30$ mK) and shielded from the environment by superconducting and magnetic shields.

The coupled transmon/nonlinear oscillator is modeled by the generalized Jaynes-Cummings-Kerr Hamiltonian ($\hbar=1$),
\begin{eqnarray}
H = \omegr \adag a + K \adag \adag a a + \sum_i \epsilon_i |i\rangle \langle i| + \nonumber\\+ \sum_{i,j=0}^{N_l} g_{ij} |i\rangle\langle j|(\adag  + a), \label{eq:1}
\end{eqnarray}
where $\epsilon_i$ are the transmon energy levels, $g_{ij}$ are the coupling matrix elements between the transmon and the cavity \cite{koch07transmon}, $N_l$ is the effective number of transmon levels, and $\adag (a)$, is the creation (annihilation) operator for the resonator mode. Terms in the Hamiltonian describing dissipation and excitation have been suppressed. Here we have modeled the large critical current junction as a Kerr nonlinearity\cite{ong11nonlinear} with Kerr coefficient $K/2\pi =-60$ kHz since the junction is inductively shunted and results only in weak anharmonicity. The transmon junction inductance, on the other hand, participates strongly in the circuit resulting in quantized anharmonicity where only a few quantum levels participate in the dynamics.  To illustrate the effect of the qubit on the resonator mode, we can write an effective Hamiltonian for the system that is valid for a small number of excitations in the dispersive regime $|\Delta|\gg g_{01}$ ($\Delta\equiv \epsilon_1-\epsilon_0-\omega_0$),
\begin{eqnarray}
H_\textrm{eff} = \epsilon+ \omega \hat{n} - \lambda \hat{n}^2, \label{eq:heff}
\end{eqnarray}
where $\hat{n}=a^\dagger a$ is the cavity photon number operator, and $\epsilon,\ \omega,$ and $\lambda$ are the effective qubit energy, resonator frequency, and nonlinearity of the system.   In the classical Duffing oscillator Hamiltonian, the AR threshold has been shown to be sensitively dependent on the nonlinearity\cite{marc04ladder}. As such, we expect the dependence of $\lambda$ on detuning and the qubit state to influence the AR threshold.

\begin{figure}
\includegraphics[angle = 0, width = 0.45\textwidth]{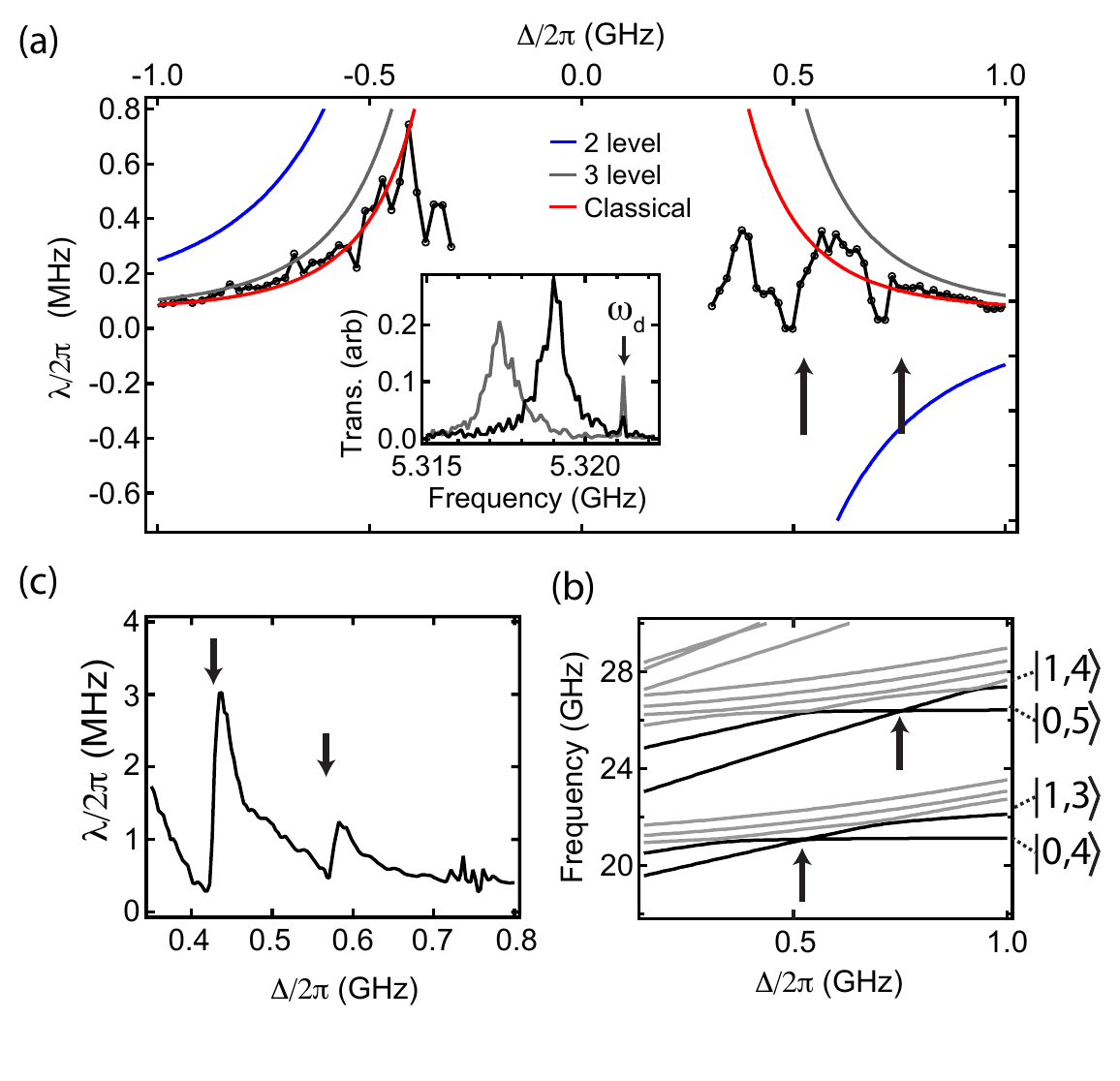}
\caption{\label{fig2} Energy levels and the effective nonlinearity $\lambda$ of the strongly coupled system.  (a) The measured coefficient of nonlinear response for a strongly coupled system versus qubit-cavity detuning for the qubit prepared in the ground state in the low excitation regime.  Theory curves for the model Eq.\ (\ref{eq:1}) with $N_l=2$ (blue), $N_l=3$ (gray), and for a model of coupled nonlinear classical oscillators (red) are shown.  The arrows indicate the locations of avoided crossings of the level pairs $|1,3\rangle \leftrightarrow |0,4\rangle$  and $|1,4\rangle \leftrightarrow |0,5\rangle$. (Inset) The transmission of the resonator when driven with tone at $\omega_\mathrm{d}$ that occupied the resonator with $\nbar = 0.4$ (black) and $\nbar = 10$ (gray) off-resonant photons.  (b) Energy levels of the qubit-oscillator model with $N_l = 7$ show the avoided crossings in the 4 and 5 excitation manifold. (c) Quantum trajectory simulation of the system  exhibits a general trends of increasing effective nonlinearity $\lambda$ with diminishing  qubit-cavity detuning ($\Delta$) with abrupt reductions associated with avoided crossings in the 4 and 5 excitation manifold. For these simulations, the qubit energy levels were modeled as a Duffing nonlinearity.}
\end{figure}

To quantify the role of $\lambda$ in determining the AR threshold, we begin by characterizing the low-power steady-state response of the system as the qubit detuning is varied to be positive or negative.  The transmission of the oscillator, shown in Figure 2, was recorded with a vector network analyzer using a weak probe tone, while the power of an off-resonant drive at frequency $\omega_\mathrm{d}$ was varied.  The number of off-resonant photons was calibrated using the ac stark shift for $\Delta/2\pi = 2.64$ GHz, and using the measured cavity Q at each detuning.  As shown in Figure 2(a), the resonant response of the nonlinear resonator shifts to lower frequency for increasing drive powers.  As a note, we also observed a reduction of the resonator Q when the qubit is near resonance.  This reduction may be attributed to a power broadening as the excitation gradually becomes more qubit-like. At each detuning, we extract the effective nonlinearity, $\lambda$, of the resonator by fitting the power-dependent resonance frequency to a straight line.  These results are plotted in Figure 2(a) as a function of the detuning.  We also plot the calculated nonlinearity for $N_l = 2$ and $N_l = 3$ transmon levels.  The calculated nonlinearity for $N_l = 2$ agrees poorly with the data, indicating that higher levels of the transmon should be included in the analysis, even in the regime of weak excitation\cite{footnote2}.  A calculation based on coupled, classical nonlinear oscillators agrees well with the data for $\Delta<0$, but does not capture the significant reductions in the nonlinearity observed for $\Delta/2\pi = +0.5$ and $\Delta/2\pi = +0.7$.  To explain the origin of these features,  we show the calculated energy levels of the system as a function of detuning, obtained by numerical diagonalization of Eq.\ (\ref{eq:1}) with 7  transmon levels and $K=0$ in Figure 2(b).  The energy levels show clear avoided crossings for $|0,n\rangle$ and $|1,n-1\rangle$ energy levels for $n=4$ and $n=5$ near $\Delta/2\pi =0.5$ and $\Delta/2\pi = 0.7$.  We note that the avoided crossings for the $n$ excitation manifold are large, on the order of $\sqrt{n}g$ and introduce gaps that effectively isolate the  lower manifold of photon excitations from the rest of the anharmonic ladder. With the probe employed at these frequencies, the classical nonlinearity is disrupted by the avoided crossings and the description of the system with a constant anharmonicity is insufficient. This effect manifests as a  reduction of the measured nonlinearity for these detunings.  To verify this, we have simulated the nonlinear response of the system when the qubit energy levels are modeled as a Duffing nonlinearity.  The energy levels exhibit the same structure, and the simulated nonlinearity exhibits corresponding abrupt reductions as displayed in Figure 2(c). 
 
\begin{figure}
\includegraphics[angle = 0, width = 0.5\textwidth]{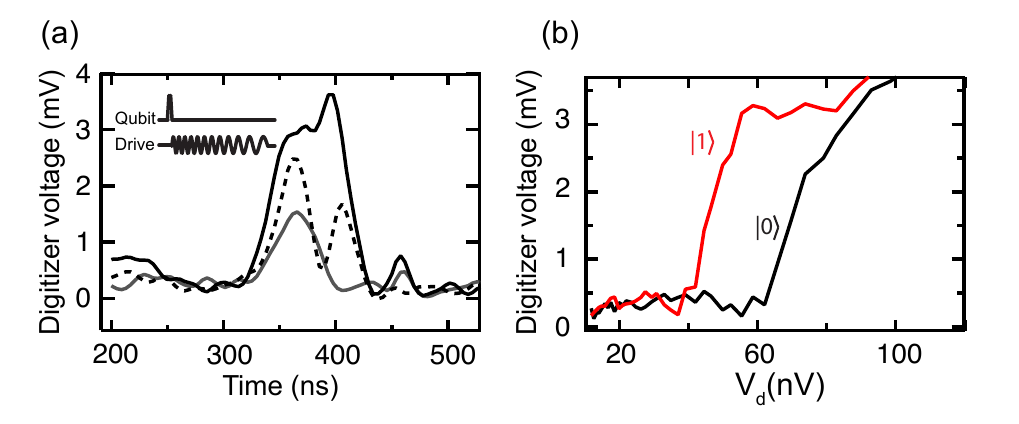}
\caption{\label{fig3duffing}    (a) The transmitted magnitude for chirp sequences with drive voltages that were above (black), near (dashed), and below (gray) the AR threshold.  (Inset) Pulse sequence:  the qubit manipulation pulse was applied immediately before the start of the chirp sequence. (b) The average transmitted magnitude near 400 ns versus drive voltage shows $S_{|0\rangle}$ (black) and $S_{|1\rangle}$ (red) for $\Delta/2\pi = 0.59$ GHz. }
\end{figure}

\begin{figure}
\includegraphics[angle = 0, width = 0.45\textwidth]{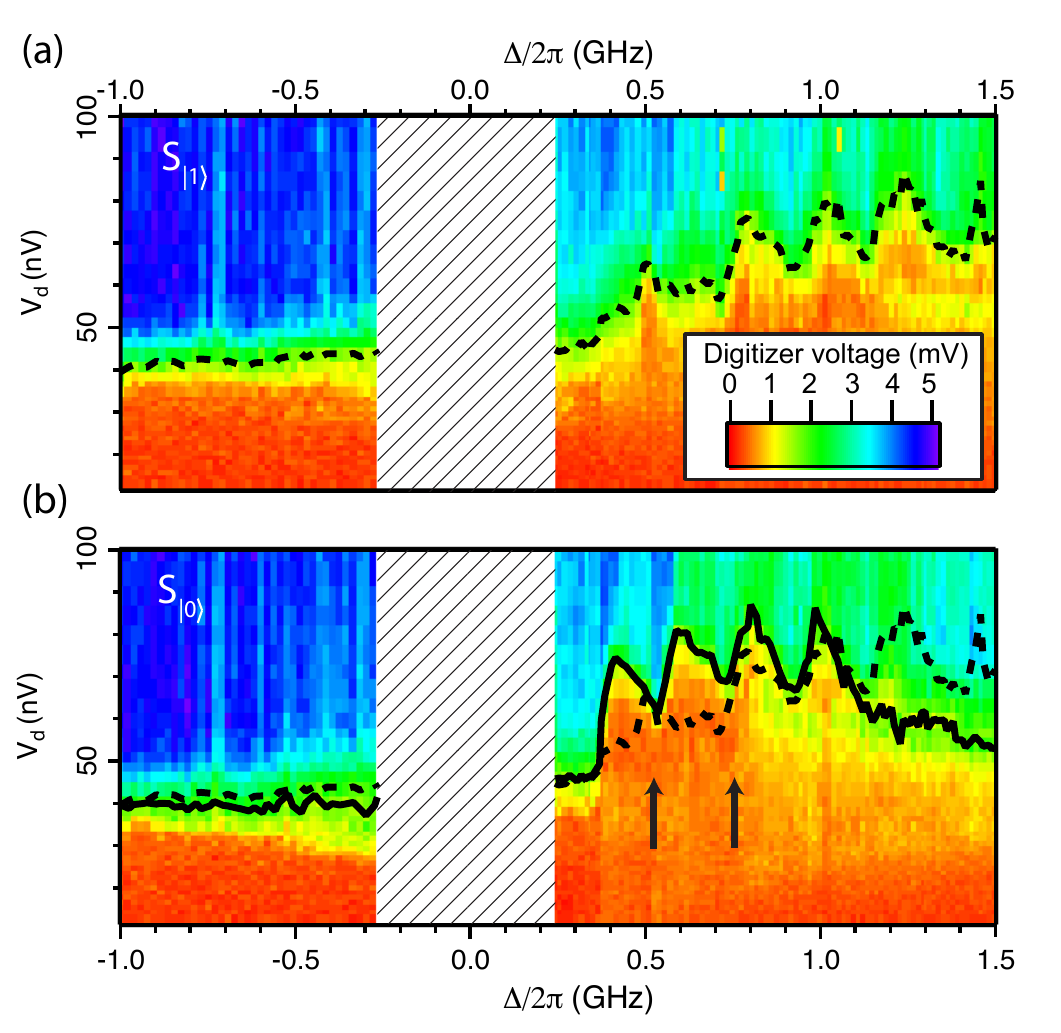}
\caption{\label{fig2}
Measured autoresonance and threshold sensitivity on the level structure and initial qubit state. (a) Color plot shows $S_{|1\rangle}$ versus qubit detuning.  The dashed line indicates the AR threshold,  $V_{|1\rangle}$.  AR measurements were not taken for small values of the detuning as indicated by the hatched region.  (b) Color plot of $S_{|0\rangle}$ with $V_{|0\rangle}$ indicated as a solid black line.  The AR threshold, $V_{|1\rangle}$, is also plotted for comparison as a dashed black line. The two arrows indicate the location of avoided crossings in the 4 and 5 excitation manifold.}
\end{figure}

For the range of detunings that we probed, the coupled system nonlinearity in the low excitation regime is dominated by the qubit, and consequently reflects the quantized levels of the system. In contrast, when $\nbar\gg 1$ a large number of excitations are added to the coupled cavity-qubit system, the effect of the qubit is diminished \cite{levprl} and the effective nonlinearity of the cavity is dominated by the weaker Kerr term.  Since the Kerr nonlinearity originates from a relatively large critical  current Josephson junction, the nonlinearity persists to high powers.  To study this regime, we now turn to examining the transient response of the system to a frequency chirped drive.  The chirp excitation pulse, generated using a voltage controlled oscillator as previously described\cite{murc11chirp}, started at $5.54$ GHz and was ramped down to $5.14$ GHz in 500 ns. In Figure 3(a) we show the response of the oscillator to the chirped excitation for varying drive amplitudes.  The data are the average of 10,000 chirp sequences.  As the drive approaches the resonance, energy is transferred and oscillation amplitude begins to build.  When the drive amplitude is weak (gray), we observe that after small excitation the system relaxes back to low amplitude oscillations.  In contrast, when the drive is above (black) a threshold value, %$V_c\propto \omega^{-1/2} \alpha^{3/4}$,
 the oscillator builds up energy and oscillates at high amplitude, behavior indicative of phase locking of the oscillations to the drive. Near the threshold, in a given chirp sequence, the resonator probabilistically locks or not into high amplitude oscillations, resulting in the dashed average trace. 

Examining the response of the resonator at a specific point in the chirp sequence, the average amplitude of oscillations shows a typical ``$S$-curve'' as shown in Figure 3(b) that transitions from a low average transmitted voltage, corresponding to no locking events to high average voltage corresponding to a locking probability $P=1$. The threshold exhibits a finite width that arises from  the  initial quantum fluctuations of the oscillator and can be explained semiclassically\cite{murc11chirp}.   To examine the effect of different quantum initial states on the AR dynamics we compare $S$-curves for the qubit prepared in either the $|0\rangle$ or $|1\rangle$ state.  The $|1\rangle$ state was prepared by applying a $\pi$ pulse to the qubit immediately before the start of the chirp sequence. In Figure 3(b) we display $S_{|0\rangle}$ and $S_{|1\rangle}$, $S$-curves for the initial $|0\rangle$ and $|1\rangle$ states for  $\Delta/2\pi= +0.59$ GHz.  We note that the AR threshold for the two states are significantly different, and correspond to a maximum discrimination or AR capture fidelity\cite{supplement} of $ 90\%$.  Given perfect distinguishability between the low- and high-amplitude responses of the oscillator, the AR capture fidelity would correspond to the single shot readout fidelity. We note that this value is consistent with near-unity fidelity for  mapping the initial qubit state onto the AR response of the oscillator after accounting for decay of the qubit $|1\rangle$ state due to a finite $T_1 = 1\ \mu$s.  

In Figure 4(a) we plot $S_{|1\rangle}$ versus qubit detuning.  The dashed line indicates $V_{|1\rangle}$, which is defined as the drive voltage where the locking probability $P = 1/2$ for the qubit prepared in the excited state. Similarly, Figure 4(b) displays $S_{|0\rangle}$, with the solid line indicating $V_{|0\rangle}$, the corresponding threshold value for the ground state.  Note that $V_{|1\rangle}$ is superimposed for comparison.  Similar to the low power nonlinearity, the response is asymmetric for positive versus negative detuning.  When the qubit is below the resonance, the threshold for the excited state is higher than that for the ground state, consistent with the expected dependence\cite{naam08chirp} of $V_c\propto \omega^{-1/2}$ based on the dispersive cavity shift due to the qubit, as well as with the dependence of $\lambda$ (Eq. \ref{eq:heff}) on the qubit state. 

In contrast, when the qubit is tuned above the cavity resonance, the threshold for the excited state is reduced compared to the ground state, which varies dramatically with qubit detuning up to $\Delta/2\pi\sim1$ GHz, where again the threshold for the qubit exited state is higher compared to the ground state. Surprisingly, the quasi-periodic structure of the high power signal emanating from thousands of photons in the cavity reflects the detailed level structure of the cavity QED model in  Eq.\ (\ref{eq:1}). For $\Delta>0$, this level structure is significantly perturbed due to many avoided crossings in the transmon-cavity spectrum. For the qubit in the ground state, the values of $\Delta$ for which the threshold is sharply reduced correspond  to the avoided crossings that cause drastic changes to the low power nonlinearity. The reductions of the threshold for the excited qubit state fit well with the next set of avoided crossings between the $|1,n\rangle$ and $|2,n-1\rangle$ levels.

In conclusion, we have demonstrated AR capture in a weakly nonlinear oscillator coupled  strongly to a quantum system. The interaction modifies the AR threshold depending on the specific qubit-oscillator level structure, and provides a sensitive probe of the initial quantum state. We have harnessed this effect to realize a high-fidelity, latching qubit readout whose speed, unlike techniques employing amplitude modulation \cite{sidd06jba, mall09, levprl,reed10,gino10,bois10}, is not limited by the oscillator Q. In our experiment, a frequency chirp pulse sequence of 200 ns is sufficient for readout whereas amplitude modulation would require a  $\sim 3\  \mu$s pulse for a our $\mathrm{Q} =9000$ resonator, far exceeding the $T_1$ lifetime of the qubit. As such, this new measurement technique is potentially a valuable resource for new types of superconducting qubit circuits which incorporate very high-Q superconducting 3D resonators\cite{hanhee_paik_highQ}.

%High-Q cavities are of  interest in decreasing the Purcell decay rate of the qubit and coupling to spurious and higher order electromagnetic modes.  However, high-Q cavities limit the rate at which information can be extracted from the system.
 %The non-equilibrium excitation of of a high-Q nonlinear resonator allows for readout of the qubit state in a time that is not limited by the cavity linewidth. 
%The ability to read out the state of a qubit coupled to a high-Q cavity at a rate that is not limited by the cavity linewidth could be especially important as new generations of very high-Q superconducting 3D resonators\cite{hanhee_paik_highQ} are of increasing interest for  applications in quantum information processing.%,

This research was supported in part by the National Science Foundation (DMR-1004406),
the U.S. Army Research Office (W911NF-11-1-0029 and W911NF-09-1-0514), the facilities and staff of the Yale University Faculty of Arts and Sciences High Performance Computing Center, 
and the Office of the Director of National Intelligence
(ODNI), Intelligence Advanced Research Projects Activity
(IARPA), through the Army Research Office. All statements of fact,
opinion or conclusions contained herein are those of the
authors and should not be construed as representing the
official views or policies of IARPA, the ODNI, or the US
Government. EG acknowledges support from EPSRC (EP/I026231/1).

%This research was funded by the Office of the Director of National Intelligence (ODNI), Intelligence Advanced Research Projects Activity (IARPA), through the Army Research Office. All statements of fact, opinion or conclusions contained herein are those of the authors and should not be construed as representing the official views or policies of IARPA, the ODNI, or the US Government. 

%   \begin{figure}
%\includegraphics[angle = 0, width = 0.5\textwidth]{figures/fig1}
%\caption{\label{fig3} Measurement setup. (a) Homodyne measurement of the phase noise.  The  sign of the probe tone is chopped at 1 kHz to subtract 1/f noise in the room temperature components. A cryogenic switch allows calibration of the 1/f noise floor.  (b) The phase response of a typical 1-port device allows calibration of the absolute frequency noise of a resonator.   (c)  Typical 1/f frequency noise spectral density of a typical lumped element resonaor. }
%\end{figure}

%\bibliographystyle{prsty}
%\bibliography{allrefs_2011_kwm}

\begin{thebibliography}{18}
\bibitem[{foo()}]{footnote2}
\bibinfo{note}{We note that higher levels of the transmon have a similar effect
  on the dispersive frequency shift, $\chi$.}
\end{thebibliography}

\hspace{-.5in}\includegraphics[angle = 0, width = 1.1\textwidth]{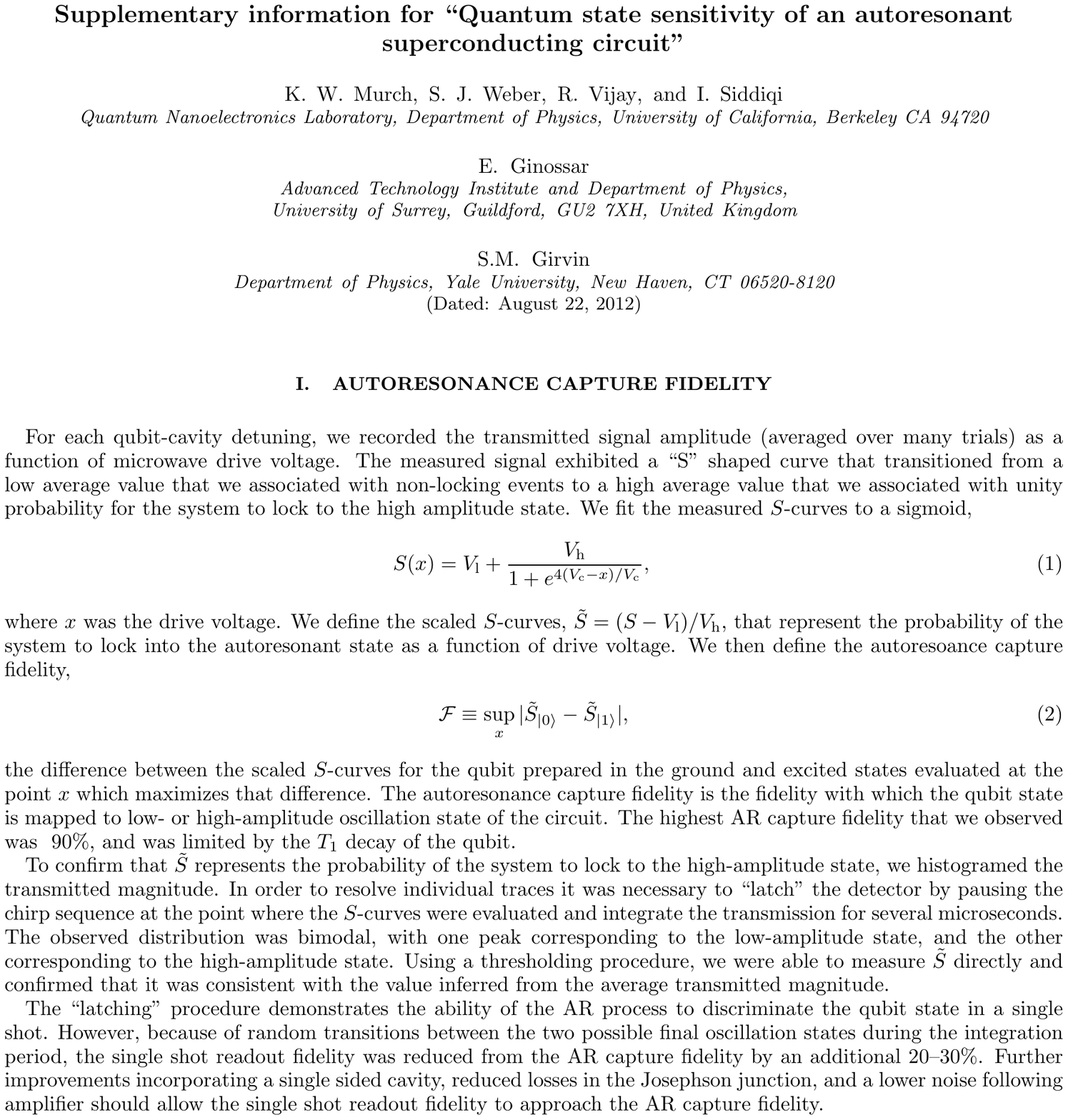}

\end{document}